

\newcount\sectionno \sectionno=0
\newcount\eqnnumber \eqnnumber=0
\newcount\subsectionno\subsectionno=1
\font\bbfr=msym10
\newfam\bbffam 
\textfont\bbffam=\bbfr

\def\cdd{{\cdot}}
\def\ref#1{${}^{[#1]}$}
\def\pis{\Pi\kern-6.5pt /\kern2pt}
\def\ps{P\kern-6.5pt /\kern2pt}
\def\os{{\omega\kern-5.5pt /\kern2pt}}
\def\xs{x \kern-5pt /\kern1pt}

\def\D{{\cal D}}

\def\om{1+2{\hbox{\raise8.05pt\vbox{\hrule width6pt}\kern-6.6pt
$\Omega$}}\Omega}
\def\omm{1+{\hbox{\raise8.05pt\vbox{\hrule width6pt}\kern-6.6pt
$\Omega$}}\Omega}

\def\low#1{\lower3pt\hbox{$\scriptstyle #1$}}
\def\overbarD{\hbox{\raise8.05pt\vbox{\hrule width6pt}\kern-7pt
$D$}}
\def\overbarZ{\hbox{\raise8.05pt\vbox{\hrule
width6pt}\kern-7.2pt $Z$}}

\catcode`\@=11
\def\qmatrix #1#2{\,\left(\null\,\vcenter{
\normalbaselines\openup#1\jot\m@th
    \ialign{\hfil$##$\hfil&&\quad\hfil$##$\hfil\crcr
      \mathstrut\crcr\noalign{\kern-\baselineskip}
      #2\crcr\mathstrut\crcr
\noalign{\kern-\baselineskip}}}\,\right)\,}
 \catcode`\@=12

\def\appendix#1{

   \eqnnumber=0
   \def\nextno{
       \global\advance\eqnnumber by 1
       ({\rm #1}\cdd\the\eqnnumber)
              }
\def\eqlabel##1{
       \xdef##1{{$({\rm #1}\cdd\the\eqnnumber)$}}
      }
}

\def\newsection#1{
   \vskip .5truecm \goodbreak
   \global\advance\sectionno by 1
   \leftline{\bf\the\sectionno\quad  #1}
   \nobreak\vskip .2cm\nobreak
   \eqnnumber=0 \subsectionno=1
\def\nextno{
       \global\advance\eqnnumber by 1
       \eqno(\the\sectionno\cdd\the\eqnnumber)
              }
\def\nextnoalign{
       \global\advance\eqnnumber by 1
       (\the\sectionno\cdd\the\eqnnumber)
              }
\def\eqlabel##1{
       \xdef##1{{$(\the\sectionno\cdd\the\eqnnumber)$}}
      }
\def\newsubsection##1{
   \vskip .5cm \goodbreak
   \global\advance\subsectionno by 1
   \eqnnumber=0
   \leftline{\bf\the\sectionno.\the\subsectionno\quad  ##1}
   \nobreak\vskip .2cm\nobreak
   \eqnnumber=0

}}

%
%
%
\newwrite\refout
\newcount\refno
\def\refbegin{\immediate\openout\refout=refout.tex\refno=1}

%
%
\def\form#1#2#3#4#5#6{\xdef#1{{\noexpand\rm#2,
 {\noexpand\rm#3\/}
  {\noexpand\bf#4} (#5) #6.}}}
%
%
\def\bookform#1#2#3#4{\xdef#1{{\noexpand\rm#2, in
{\noexpand\it#3\/}\hfil
  \break\line{\hfill(#4).\qquad}}}}
%
%
\def\procform#1#2#3#4#5#6{\xdef#1{{\noexpand\rm#2, in
{\noexpand\it#3\/}
  (#4), #5 (#6).}}}
%
%
\def\prepform#1#2#3#4#5{\xdef#1{{\noexpand\rm#2, #3 (#4)#5.}}}
%
%
\def\freeform#1#2{\xdef#1{{\noexpand\rm#2.}}}
%
%
\def\refwrite#1{\immediate\write\refout{\noexpand#1}}
\def\ref#1#2{\ifx#1\undefined\message{*Ref*}$\ast$\else
  \ifx#2\undefined\xdef#2{\number\refno}#2%
  \refwrite{\item{[#2]}\noexpand#1}%
  \global\advance\refno by 1\else#2\fi\fi}
%
%
\def\xref#1#2{\ifx#1\undefined\message{*Ref*}$\ast$\else
  \ifx#2\undefined\xdef#2{\number\refno}%
  \refwrite{\item{[#2]}\noexpand#1}%
  \global\advance\refno by 1\else\fi\fi}
%
%
\def\sref#1{\ifx#1\undefined\message{*Ref*}$\ast$\else
  \refwrite{\item{}\noexpand#1}\fi}


\form{\arms}{J.\ Arms}{J.\ Math.\ Phys.}{18}{1977}{830}

\form{\armstwo}{J.\ Arms}{J.\ Math.\ Phys.}{20}{1979}{443}

\form{\armsmars}{J.\ Arms, J.\ Marsden and V.\ Moncrief}
{Ann.\ Phys.}{144}{1982}{81}

\form{\armsmarstwo}{J.\ Arms, J.\ Marsden and V.\ Moncrief}
{Commun.\ Math.\ Phys.}{78}{1981}{455}

\form{\brill}{D.\ Brill and S.\ Deser}{Comm.\ Math.\ Phys.}
{32}{1973}{291}

\freeform{\fischer}{A.\ E.\ Fischer, in {\it{Relativity}},
eds. M.\ Carmeli, S.\ I.\ Fickler and L. Witten, (Plenum,
New York, 1970)}

\form{\fischeretal}{A.\ Fischer, J.\ Marsden and V.\ Moncrief}
{Ann.\ Inst.\ Henri\ Poincar\' e}{33}{1981}{147}

\freeform{\fischmars}{A.E.\ Fischer and J.E.\ Marsden, in
{\it{General Relativity: An Einstein Centenary Survey}}, eds.
S.W.\
Hawking and W.\ Israel, (C.U.P., Cambridge, 1979)}

\form{\hig}{A. Higuchi}{Class.\ Quant.\ Grav.}{8}{1991}{1961}

\form{\higtwo}{A. Higuchi}{Class.\ Quant.\ Grav.}{8}{1991}{1983}

\form{\higthree}{A. Higuchi}{Class.\ Quant.\ Grav.}{8}{1991}{2005}

\form{\moncrief}{V.\ Moncrief}{J.\ Math.\ Phys.}{16}{1975}{493}

\form{\moncrieftwo}{V.\ Moncrief}{J.\ Math.\ Phys.}{17}{1976}{1893}

\form{\moncriefthree}{V.\ Moncrief}{Phys.\ Rev.}{D13}{1978}{983}

\freeform{\wald}{Private Communication}

\form{\waldsud}{D.\ Sudarsky and R.M.\ Wald}{Phys. Rev.}{D46}
{1992}{1453}

\form{\lifschitz}{E.M.\ Lifschitz and I.M.\
Khalatnikov}{Adv.\ Phys.}{12}{1963}{185}

\form{\hallhawk}{J.J.\ Halliwell and S.W.\
Hawking}{Phys.\ Rev.}{D31}{1985}{1777}

\freeform{\fano}{U.\ Fano and G.\ Racah, {\it Irreducible
Tensorial Sets} (Academic Press, New York, 1959)}

\freeform{\dowker}{J.S.\ Dowker and D.\ Pettengill, Dept.
of Physics, University of Manchester report (1974)
unpublished}

\freeform{\edmonds}{A.R.\ Edmonds, {\it Angular Momentum
in Quantum Mechanics} (Princeton University Press, 1957) p.95,
130-132}

\form{\dowlaf}{H.F.\ Dowker and R.\ Laflamme}{Nucl.\ Phys.}
{B366}{1991}{209}

\freeform{\traskast}{D.A.\ Kastor and J.\ Traschen,
``Linear Instability of Non-vacuum
Spacetimes,'' SLAC-PUB-5664 (1991)}


\refbegin
\def\half{{1\over 2}}
\def\H{{\cal{H}}}
\def\D{{\cal{D}}}

\magnification=\magstep 1
\openup 2 \jot

{\rightline{FERMILAB-Pub-92/294-A}}

{\rightline{ October 1992}}

\vskip .5cm
{\centerline{ SO(4) INVARIANT STATES IN QUANTUM COSMOLOGY}}

\vskip .5cm
{\centerline{H.F. Dowker}}
\vskip .3cm
{\centerline{NASA/Fermilab Astrophysics Center}}
{\centerline{Fermi National Accelerator Laboratory}}
{\centerline{P.O. Box 500, Batavia, IL 60510, U.S.A.}}
\vskip .3cm

{\centerline{E-mail address: dowker@fnas10.fnal.gov}}
\vskip 1cm
{\centerline{\bf{ABSTRACT}}}
 The phenomenon of linearisation instability is
identified in models of quantum cosmology
that are perturbations of mini-superspace models.
In particular, constraints that are second order
in the perturbations must be imposed on wave functions
calculated in such models. It is shown explicitly that
in the case of a model which is a perturbation of
the mini-superspace which has $S^3$ spatial sections
these constraints imply that any wave functions calculated
in this model must be $SO(4)$ invariant.
\vfill\eject

\newsection{Introduction}

The phenomenon of linearisation instability in
classical general relativity is well understood
[\ref\brill\BRILL- \xref\fischmars\FISCHMARS
\ref\moncrief\MONCRIEF].
It arises when approximations to solutions of
the vacuum Einstein equations are sought
by expanding the equations about a known solution
 which has compact Cauchy surfaces and non-trivial
Killing
vectors and solving the linearised equations for
the perturbation. In this case, solving the
linearised equations alone does not always yield
a metric which is a good approximation to a solution
of Einstein's equations i.e. a solution to the
linearised equations may not be tangent to a curve
of exact solutions.

The reason is that some of the
constraints of general relativity are exactly zero to
linearised order, in fact there is one such
constraint  for every Killing vector. Thus the first
non-zero order is the second and there is one second
order constraint for each Killing vector. Imposition
of these second order constraints is what is needed
to eliminate the spurious solutions. These complications
can also be seen as a reflection of the structure of the
space of solutions that are close
to a solution with  Killing
vectors and compact Cauchy surfaces
[\ref\fischeretal\FISCHERETAL
\xref\armsmars\ARMSMARS-\ref\armsmarstwo\ARMSMARSTWO].
This space is not a manifold, since the diffeomorphism
group does not act freely but has a fixed point which
is precisely the background
metric with isometries.
Rather, it has a stratified stucture and
the background geometry is a singular point in the space.

It has been pointed out that when one comes to quantise
gravitational perturbations on backgrounds with compact
Cauchy surfaces and Killing vectors, one must again
take into consideration these second order constraints,
now imposed as operators  annihilating physical states.
[\ref\moncriefthree\MONCRIEFTHREE].
The consequences of this have been worked out in
detail for the case of DeSitter space [\ref\hig\HIG \sref\higtwo
\sref\higthree].

Although linearisation instability would not be expected to
play a role in quantum cosmology in
general since one integrates over all
four-geometries, symmetries or
not,  it does turn out to be important in
models of quantum cosmology in which departures from
mini-superspace
are considered small in some sense. In these cases, the
mini-superspace has closed (compact without boundary) spatial
sections  and spatial Killing vectors   and the same
considerations
as before must be made.

The purpose of the present paper is to demonstrate how
linearisation
instability arises in the quantum cosmology model of
Halliwell and Hawking [\ref\hallhawk\HALLHAWK].
The paper proceeds as follows. In Section 2, we review
classical linearisation instability.
In Section 3  the model is described
and we see explicitly how six of the linearised momentum
constraints vanish identically. The expressions
for the second order constraints are derived.
Section 4 contains the calculation of the quantum second
order constraints in a representation on wave functions that
are functions of the scale factor and the mode coefficients
of the harmonic expansion of the perturbation.
It is shown that
these six constraints obey the algebra of $SO(4)$.
In section 5 it is shown
In section 6  a scalar field is added  the
analysis repeated. Section 7 is a discussion.

\newsection{Linearisation Instability}

This brief discussion follows that of Moncrief
[\ref\moncriefthree\MONCRIEFTHREE].
Let M be a compact three-manifold without boundary.
In the hamiltonian formulation of general relativity,
the dynamical variables are $(g,\pi)$, where $g=g_{ij}$
is a riemannian metric and $\pi=\pi^{ij}$ is its
canonical momentum, a tensor density, on M. Due to
diffeomorphism invariance, general relativity is a
constrained theory. The constraint hypersurface in
phase space is defined by $\Phi(g,\pi)=0$, where
$\Phi$ is the constraint map
$$
\Phi(g,\pi) = \left(\H(g,\pi), \H^i(g,\pi)\right),
\nextno
$$
with
$$  \eqalignno{
{\cal{H}}(g,\pi)
&\equiv (\mu_g)^{-1}\half \left(g_{ik}g_{jl}+
g_{il}g_{jk}-g_{ij}g_{kl}\right)\pi^{ij}\pi^{kl}
-\mu_g(R(g)-2\Lambda),
&\nextnoalign\cr
\H^i(g,\pi)
&\equiv -2 \pi^{ij}{}_{;j}.
&\nextnoalign\cr
}
$$
Here, $\mu_g=({\rm{det}}g)^\half$, semi-colon denotes
covariant derivative with respect to $g$, and units have been
chosen
in which $16\pi G=1$.

Let $(g_0,\pi_0)$ be a solution of the constraints.
Suppose we are looking for a solution of Einstein's
equations close to a background solution, ${}^4g_0$, for which
$(g_0,\pi_0)$ is the initial data. The new solution will have
initial
data $(g,\pi)=(g_0+h, \pi_0+\omega)$. One may expand out the
constraints:
$$
\Phi(g_0+h, \pi_0+\omega)
=\Phi^{(1)}_{(g_0, \pi_0)}(h,\omega)
+\Phi^{(2)}_{(g_0, \pi_0)}(h,\omega) + \dots
\nextno
$$
where $\Phi^{(1)}$ ($\Phi^{(2)}$ etc.) is linear (quadratic etc.)
in the perturbation
$(h,\omega)$.
We will adopt similar notation from here on, so that a superscript
$(k)$ denotes a quantity that is $k$th
order in  $(h,\omega)$.
The usual linear constraints are $\Phi_1(g_0,\pi_0)=0$.
However, if ${}^4g_0$ admits Killing vectors, imposing the linear
constraints alone will not in general exclude nonintegrable
perturbations.

To see this, let $C$ be any function and $Y=Y^i$ be any vector
field on
M. Define the projection, $P_{(C,Y)}(\Phi)$, of $\Phi$ along
$(C,Y)$ by
$$
 \eqalignno{P_{(C,Y)}\left(\Phi\right) &=
\int_M d^3x \left<(C,Y), \Phi(g,\pi)\right>\cr
&=\int_M d^3x\left[C\H(g,\pi)
+Y^i\H_i(g,\pi)\right],&\nextnoalign\eqlabel\proj\cr
}
$$
where $\H_i=g_{ij}\H^j$.
For any $(C,Y)$
$$P_{(C,Y)}\left(\Phi(g,\pi)\right) =
P^{(1)}_{(C,Y)}\left(\Phi(g,\pi)\right)
+P^{(2)}_{(C,Y)}\left(\Phi(g,\pi)\right)+\dots =0.\nextno
$$
where
$$P^{(k)}_{(C,Y)}\left(\Phi(g,\pi)\right)=
\int_M \left<(C,Y),\Phi^{(k)}_{(g_0,\pi_0)}(h,\omega)
\right>,\ k=1,2\dots. \nextno
$$
It can be shown that $P^{(1)}_{(C,Y)}\left(\Phi(g,\pi)\right)$
vanishes if and only if $C$ and $Y$ are the normal and
tangential projections on the initial surface of a Killing vector
of ${}^4g_0$. In that case, the lowest non-trivial order for the
constraint projected along the Killing direction is the second.
Thus, in order to treat the constraints consistently,
one must impose
$$\Phi_1(g_0,\pi_0)=0,\nextno\eqlabel\lincon
$$
the usual linear constraints  and, in addition,
$$
P^{(2)}_{(C,Y)}\left(\Phi(g_0+h,\pi_0+\omega)\right)=0
\nextno\eqlabel\quadcon
$$
for each Killing vector $(C,Y)$ of the background.

On quantisation of the perturbations on the background ${}^4g_0$,
 \lincon\
and \quadcon\ can be implemented as operator constraints on
physical states.

\newsection{Perturbed Mini-Superspace}

In general, one would not expect the phenomenon of linearisation
instability to arise in quantum gravity since, roughly, one integrates
over all four-geometries, with no restriction on symmetry properties.
However, in quantum cosmology, motivated by the approximate
homogeneity of the observed universe, models have been
studied in which
one imposes severe symmetries on the four geometries included in the
path integral. Going beyond these ``mini-superspace'' models,
attempts have been made to treat departures from homogeneity
perturbatively.
When the homogeneous ``background'' space is a three-sphere,
linearisation
instability emerges as expected. In this section we describe
just such a model [\ref\hallhawk\HALLHAWK].
The
three-metric, $g_{ij}$ has the form
$$
g_{ij}=a^2(t)\left(q_{ij}+h_{ij}\right)
\nextno
$$
where $q_{ij}$ is the round metric on $S^3$,
normalised so that
$\int \sqrt q  d^3x =16\pi^2$ (note that $q_{ij}=4\Omega_{ij}$
where $\Omega_{ij}$ is the metric induced by the embedding
of $S^3$ in $R^4$). $h_{ij}$ is a perturbation and to be considered as
small.

There are six Killing vectors of the homogeneous background: the three
left invariant plus the three right invariant vector fields on
$S^3$, $\{e_A{}^i: A=1,2,3\}$ and  $\{\tilde e_A{}^i\}$,
respectively. They
satisfy
$$
[e_A,e_B]=-\epsilon_{AB}{}^C e_C,\quad
[\tilde e_A,\tilde e_B]=\epsilon_{AB}{}^C \tilde e_C,\quad
\delta^{AB}e_A{}^i e_B{}^j=\delta^{AB}\tilde e_A{}^i e_B{}^j =q^{ij}.
\nextno
$$
They are, in fact, the lowest order vector harmonics on $S^3$
(Lifschitz harmonics $(S^{o,e}_{n=1})^i$ [\ref\lifschitz\LIFSCHITZ]).
We introduce an alternative, ``spherical'',
 basis for the Killing vectors,
$\{e_a{}^i: a=\pm 1, 0\}$ defined by
$$ \eqalignno{
e_{\pm 1}&=\pm {1\over{\sqrt 2}}(e_1\mp ie_2),\cr
e_0&=-ie_3,&\nextnoalign\cr}
$$
and the dual basis of one forms,
$\{e^a{}_i\}$ such that $e^a{}_i e_b{}^i = \delta^a{}_b$.
$\{\tilde e_a{}^i\}$
and $\{\tilde e^a{}_i\}$ are defined similarly.

Let us rename the projected constraints,
$P_{(0,e_a)}(\Phi)$ as $P_a$ and $P_{(0,\tilde e_a)}(\Phi)$ as
$\tilde P_a$
and expand them out in the perturbation. First consider $P_a$.
$$
\eqalignno{
P_a&=-2\int d^3x \pi^{ij}{}_{;j} e_a{}^k g_{ik}\cr
&=-2a^2\int d^3x \pi^{ij}{}_{;j} e_a{}^k (q_{ik}+h_{ik}).
&\nextnoalign
\cr}
$$
The zeroth order constraint is zero since that relates to the
background which is homogeneous: $(\pi^{ij}{}_{;j}){}^{(0)}=0$. The
first order constraint is
$$\eqalignno{
\left(\pi^{ij}{}_{;j}\right)^{(1)}=&
\pi^{(1)}{}^{ij}{}_{|j} +  \pi^{(0)}{}^{kj}
\Gamma^{(1)}{}^i_{kj}  \cr
=&
\pi^{(1)}{}^{ij}{}_{|j} +
\pi^{(0)}{}^{kj}
h^i{}_{k|j}-\half \pi^{(0)}{}^{ij}
h^k{}_{k|j}&\nextnoalign\cr
}
$$
where vertical bar denotes covariant derivative with respect to
$q_{ij}$, all tensor indices are (now and henceforth)
 raised and lowered with $q$ and
$\Gamma^i_{jk}$ is the Christoffel symbol of the metric $g_{ij}$.

So
$$\eqalignno{
P^{(1)}_a &= -2a^2
\int d^3x (\pi^{ij}{}_{;j})^{(1)} e_a{}^k q_{ik}\cr
&=-2a^2 \int d^3x \left[\pi^{(1)}{}^{ij}{}_{|j} +
\pi^{(0)}{}^{kj}
h^i{}_{k|j}-\half \pi^{(0)}{}^{ij}
h^k{}_{k|j}\right] e_a{}^l q_{il}. &\nextnoalign  \cr
}$$
Using $e_a{}^{(i|j)}=0$ and $\pi^{(0)}{}^{ij}\propto q^{ij}$, it can
be shown that
$
P^{(1)}_a=0
$
as expected and similarly
$
\tilde P^{(1)}_a=0.
$

Now let us consider the second order,
$$
P^{(2)}_a= -2a^2 \int d^3x\left[
(\pi^{ij}{}_{;j}){}^{(2)} e_a{}^k q_{ik}
+ (\pi^{ij}{}_{;j}){}^{(1)} e_a{}^k h_{ik}\right].
\nextno$$
We have
$$
(\pi^{ij}{}_{;j}){}^{(2)}
= \pi^{(2)}{}^{ij}{}_{|j} +  \pi^{(1)}{}^{kj}
\Gamma^{(1)}{}^i_{kj}  +   +  \pi^{(0)}{}^{kj}
\Gamma^{(2)}{}^i_{kj}  ,
\nextno
$$
whence
$$P^{(2)}_a= -2a^2 \int d^3x\biggl[
-\half h_{kj|i}\pi^{(1)}{}^{kj} e_a{}^i -h_{ik}\pi^{(1)}{}^{jk}
 e_a{}^i{}_{|j}
\biggr].\nextno\eqlabel\weak
$$

\newsection{The Algebra of the Second Order Constraints}

In this section we will expand $h_{ij}$ and $\pi^{ij}$ in spin-2
hyperspherical
spinor harmonics on $S^3$ (more details of which can be found in
[\ref\dowker\DOWKER,\ref\dowlaf\DOWLAF]),
and calculate the second order constraints.

$$\eqalignno{
h_{ij}(\gamma,t)=& e^a{}_i(\gamma) e^b{}_j(\gamma)
\pmatrix{1&1&m\cr a&b&2\cr} \sum_{LJ}
 h^N_L{}^M_J(t) Y^2_m{}^L_N{}^J_M (\gamma)\cr
&+ q_{ij}(\gamma)\sum_{J}x^{NM}_J(t)\D^J{}_{NM} (\gamma)
\sqrt{{2J+1}\over{16\pi^2}} ,&\nextnoalign\cr
\pi^{ij}(\gamma,t)=&{1\over {48 \pi^2}}
q^{ij}(\gamma) \sqrt q \biggl[(2a)^{-1}\pi_a
- a^{-2}\sum_{LJ} h^N_L{}^M_J(t)\pi_h{}^L_N{}^J_M(t)\cr
& -a^{-2}\sum_{J} x^{NM}_J(t)\pi_x{}_{NM}^J(t)\biggr]\cr
& +{1\over3}a^{-2}q^{ij}\sqrt q
\sum_{J} \pi_x{}^{NM}_J(t) \D^J{}_{NM}(\gamma)
\sqrt{{2J+1}\over{16\pi^2}} \cr
&+ 5a^{-2} \sqrt q e_a{}^i(\gamma) e_b{}^j(\gamma)
\pmatrix{a&b&m\cr 1&1&2\cr}\sum_{LJ}\pi_h{}^N_L{}^M_J(t)
 Y^2_m{}^L_N{}^J_M (\gamma)
 &\nextnoalign\cr
}
$$
where $\pmatrix{1&1&m\cr a&b&2\cr}$ is a three-j symbol (with the
spin-2 index $m$ raised, see eq.(4.6))
 and $
Y^2_m{}^L_N{}^J_M (\gamma)$ is a spin-2 hyperspherical spinor harmonic
on $S^3$, the point of $S^3$ being written as an element, $\gamma$,
of $SU(2)$.

$$
Y^2_m{}^L_N{}^J_M (\gamma)=\sqrt{{{(2J+1)(2L+1)}\over {16\pi^2}}}
{\cal {D}}^L{}_N{}^{N'}(\gamma) \pmatrix{L&J&2\cr{N'}&M&m\cr}
\nextno\eqlabel\harm
$$
where
$\D^L{}_N{}^{N'}(\gamma)$ is a spin-$L$ representation matrix of
$SU(2)$. From \harm\ we see there is a condition on $L$ and $J$
namely $|L-J|\leq 2$. The harmonics with $L=J$ correspond to
the ``scalar'' traceless tensor harmonics of Lifschitz, $P^n_{ij}$,
those with $|L-J|=1$ to the ``vector'' traceless tensor harmonics,
$S^n_{ij}$ and those with $|L-J|=2$ to the tranverse traceless
harmonics $G^n_{ij}$ with $n=L+J+1$ in each case.

 The harmonics are normalised so that
$$
\int d^3 x \sqrt q\;\; Y^2_m{}^L_N{}^J_M
Y{}^m_2{}^{L'}_{N'}{}^{J'}_{M'}
=\delta_{LL'}\delta_{JJ'}C^L_{NN'}C^J_{MM'}
\nextno
 $$
where
$$C^L_{NN'}=C^L{}^{NN'} =(-1)^{L-N} \delta_{N,-N'}\nextno
$$
is the spin-$L$ metric with which all
spin-$L$ indices are lowered and raised according to
$$U_N=C_{NN'}U^{N'}\quad {\rm{and}}\quad
V^N=V_{N'}C^{N'N}.\nextno
$$
Repeated indices, one upstairs and
one downstairs, are summed over.

The expansion coefficients of $\pi^{ij}$ are found using:
$$\eqalignno{
\pi_x{}^L_N{}^{}_M&\equiv
{{\partial {\cal{L}}}\over{ \partial \dot x^N_L{}^M_{}}}\cr
&=\int d^3x  {{\delta{\cal{L}}}\over{\delta \dot g_{ij}(x)}}
{{\partial \dot g_{ij} (x)}\over {\partial \dot x^N_L{}^M_{}}}\cr
&=\int d^3x \pi^{ij}(x) {{\partial \dot g_{ij} (x)}\over
{\partial \dot x^N_L{}^M}},&\nextnoalign\cr
\pi_h{}^L_N{}^J_M&\equiv
{{\partial {\cal{L}}}\over{ \partial \dot h^N_L{}^M_J}}\cr
&=\int d^3x \pi^{ij}(x) {{\partial \dot g_{ij} (x)}\over
{\partial \dot h^N_L{}^M_J}},&\nextnoalign\cr
\pi_a&\equiv
{{\partial {\cal{L}}}\over{ \partial \dot a}}\cr
&= \int d^3x \pi^{ij}(x) {{\partial \dot g_{ij}(x)}
\over{\partial \dot a}}.&\nextnoalign\cr
}$$
Note that the harmonics $\D$ and $Y$ are complex and
reality conditions on the expansion coefficients are needed
to ensure that $h_{ij}$ and $\pi^{ij}$ are real. The conditions are
that the complex conjugate of any coefficient with both indices
upstairs equals that
coefficient with its indices lowered e.g. $(h^N_L{}^M_J){}^*
=h^L_N{}^J_M$.

Substituting the expansions into \weak\ and using
$$
e^a{}_{i|j} e_b{}^i e_c{}^j = \half \sqrt 6
\pmatrix{a&1&1\cr 1&b&c\cr},\nextno
$$
$$
\D^J{}_N{}^M{}^{|i}e^a{}_i = (-1)^{2J}
\sqrt{J(J+1)(2J+1)} \D^J{}_N{}^{N'}
\pmatrix{J&M&a\cr {N'}&J&1\cr}
\nextno
$$
and the angular momentum recoupling formula
$$
\pmatrix{j_1&l_2&\mu_3\cr m_1&\mu_2&l_3\cr}
\pmatrix{\mu_1&j_2&l_3\cr l_1&m_2&\mu_3\cr}
\pmatrix{l_1&\mu_2&j_3\cr \mu_1&l_2&m_3\cr}
=\pmatrix{j_1&j_2&j_3\cr m_1&m_2&m_3\cr}
\left\{\matrix{j_1&j_2&j_3\cr l_1&l_2&l_3\cr}\right\}
\nextno
$$
(see e.g. Edmonds [\ref\edmonds\EDMONDS])
we obtain
$$\eqalignno{
P_a^{(2)}
= & \sum_{LJ} \pmatrix{1&M&J\cr a&J&{M'}\cr}
\sqrt{J(J+1)(2J+1)}\; h^{N}_L{}^{M'}_J \pi_h^{(1)}{}^L_N{}^J_M\cr
&+\sum_{J} \pmatrix{1&M&J\cr a&J&{M'}\cr}
\sqrt{J(J+1)(2J+1)}\;
 x^{N}_J{}^{M'}\pi_x^{(1)}{}^J_N{}^{}_M.
&\nextnoalign\eqlabel\fun \cr
}
$$

One way to deal with the constraints on quantisation is to
take wave functions to be functions of the coefficients,
$\Psi\equiv \Psi(a,h^N_L{}^M_J, x^N_J{}^M_{})$ and
to enforce the constraints as conditions on physical states,
representing $\pi_*$ by $-i {\partial \over{\partial *}}$. Here
we make the approximation of representing
$$\eqalignno{
\pi_x^{(1)}{}^J_N{}^{}_M&\rightarrow
-i {\partial \over{\partial {x_J^N{}_{}^M}}}\cr
\pi_h^{(1)}{}^L_N{}^J_M&\rightarrow
-i {\partial \over{\partial h^N_L{}^M_J}}.&\nextnoalign
\cr}
$$
We have
$$ \pmatrix{1&M&J\cr a&J&{M'}\cr}
\sqrt{J(J+1)(2J+1)} =-i \left(j^J_a\right)_{M'}{}^M\nextno
$$
where $\left(j^J_a\right)$ are the matrix generators of the
spin-J representation of $SU(2)$.
Thus,
 imposing the second order constraint, \fun,
gives us finally
$$
-\left[\sum (j^J_a)_{M'}{}^M \delta_{N'}{}^N h^{N'}_L{}^{M'}_J
 {\partial \over{\partial h^N_L{}^M_J}}
+\sum (j^J_a)_{M'}{}^M
\delta_{N'}{}^N x^{N'}_J{}^{M'}_{}
 {\partial \over{\partial x^N_J{}^M_{}}}\right]\Psi = 0.
\nextno\eqlabel\bag
$$

Similarly we can show that
$$
\eqalignno{
\tilde P_a^{(2)}
= & -\sum_{LJ} \pmatrix{1&N&L\cr a&L&{N'}\cr}
\sqrt{L(L+1)(2L+1)} h^{N'}_L{}^{M}_J \pi_h^{(1)}{}^L_N{}^J_M\cr
&-\sum_{J} \pmatrix{1&N&J\cr a&J&{N'}\cr}
\sqrt{J(J+1)(2J+1)}
 x^{N'}_J{}^{M}\pi_x^{(1)}{}^J_N{}^{}_M
&\nextnoalign\cr
}$$
 and thus the full set of second order quantum constraints consists of
\bag\ together with
$$
\left[\sum (j^L_a)_{N'}{}^N \delta_{M'}{}^M h^{N'}_L{}^{M'}_J
 {\partial \over{\partial h^N_L{}^M_J}}
+\sum (j^J_a)_{N'}{}^N
\delta_{M'}{}^M x^{N'}_J{}^{M'}_{}
 {\partial \over{\partial x^N_J{}^M_{}}}\right]\Psi = 0.
\nextno\eqlabel\bog
$$

\bag\ and \bog\ imply that $\Psi$ is $SO(4)$
invariant since it is easy to see that
$$\eqalignno{
\left[\hat P^{(2)}_A,\hat P^{(2)}_B\right]\Psi&=-\epsilon_{AB}{}^C
\hat P^{(2)}_C\Psi &\nextnoalign\eqlabel\ter\cr
\left[\hat{\tilde P}^{(2)}_A,\hat{\tilde P}^{(2)}_B\right]\Psi
&=\epsilon_{AB}{}^C
\hat{\tilde P}^{(2)}_C\Psi &\nextnoalign\eqlabel\tor\cr
\left[\hat{P}^{(2)}_A,\hat{\tilde P}^{(2)}_B\right]\Psi&=0
 &\nextnoalign\eqlabel\tar\cr
}$$
where  hats denote quantum
operators.
Thus the constraints generate the algebra of $SO(4)\cong
 SU(2)\times SU(2)/{Z_2}$.

Another way to see that \bag\ and \bog\ mean that $\Psi$ is $SO(4)$
invariant
 is to note that under a rotation $\gamma\rightarrow \xi\gamma
\eta^{-1}$, with $\xi,\ \eta \in SU(2)$, the coefficients
$x^{NM}_J$ and $h^N_L{}^M_J$ transform as
$$
\eqalignno{
h^N_L{}^M_J &\rightarrow {h'}{}^N_L{}^M_J =h^{N'}_L{}^{M'}_J
\D^L{}_{N'}{}^{N}(\xi^{-1}) \D^J{}_{M'}{}^{M}(\eta^{-1}) \cr
x^N_J{}^M_{}&\rightarrow {x'}{}^N_J{}^M_{} =x^{N'}_J{}^{M'}_{}
\D^J{}_{N'}{}^{N}(\xi^{-1}) \D^J{}_{M'}{}^{M}(\eta^{-1})
&\nextnoalign \cr
}$$
If $\xi$ and $\eta$ are infinitesimal we have
$$\eqalignno{
\delta h^N_L{}^M_J &=-i \left(\xi^A(j^L_A)_{N'}{}^N\delta_{M'}{}^M
+\eta^A(j^J_A)_{M'}{}^M\delta_{N'}{}^N\right) h^{N'}_L{}^{M'}_J
&\nextnoalign\cr
\delta x^N_J{}^M_{} &=-i \left(\xi^A(j^J_A)_{N'}{}^N\delta_{M'}{}^M
+\eta^A(j^J_A)_{M'}{}^M\delta_{N'}{}^N\right) x^{N'}_J{}^{M'}_{}
&\nextnoalign\cr}
$$
where $\{\xi^A\}$ and $\{\eta^A\}$ are two sets of three real parameters.
$\Psi(x,h)$ is invariant under all rotations iff
$$\left[\sum \delta h ^N_L{}^M_J
{\partial \over{\partial h^N_L{}^M_J}}
+\sum \delta x^N_J{}^M_{}
 {\partial \over{\partial x^N_J{}^M_{}}}
\right]\Psi=0 \quad \forall\ \xi^A,\eta^A
\nextno
$$
which conditions are exactly \bag\ and \bog.

We note that $\Psi\equiv \Psi (h^2, x^2)$, where
$h^2=h^N_L{}^M_J h^L_N{}^J_M$ and $x^2=x^N_J{}^M_{}
x^J_N{}^{}_M$, is $SO(4)$ invariant. More generally, a wave function
is invariant if all the ``left'' indices (i.e. indices that
transform under $\xi$) are contracted together with metrics
and/or three-j symbols and similarly for all the ``right'' indices
(that transform under $\eta$).

\newsection{The Physical Degrees of Freedom}

We are used to identifying the transverse traceless modes
of the perturbation of the gravitational field as the
physical degrees of freedom. In this section we will see that the
second order constraints can be reduced to a form that reflects
this.

We can write the constraint \fun\ as a sum of ``scalar'', ``vector''
and ``tensor'' (transverse traceless) parts
$$P_a^{(2)}={}^sP_a +{}^vP_a +{}^tP_a
\nextno
$$
where
$$
{}^sP_a = \sum_J \sqrt{J(J+1)(2J+1)}
\pmatrix{1&M&J\cr a&J&{M'}\cr}
\left(x^N_J{}^{M'}_{} \pi^{(1)}_x{}^J_N{}^{}_M
+h^N_J{}^{M'}_J \pi^{(1)}_h{}^J_N{}^J_M\right)
\nextno$$
$${}^vP_a =  \sum_J \sqrt{J(J+1)(2J+1)}
\pmatrix{1&M&J\cr a&J&{M'}\cr}
\left(h^N_{J-1}{}^{M'}_J \pi^{(1)}_h{}^{J-1}_N{}^J_{}
+h^N_{J+1}{}^{M'}_J \pi^{(1)}_h{}^{J+1}_N{}^J_M\right)
\nextno$$
$$
{}^tP_a =  \sum_J \sqrt{J(J+1)(2J+1)}
\pmatrix{1&M&J\cr a&J&{M'}\cr}
\left(h^N_{J-2}{}^{M'}_J \pi^{(1)}_h{}^{J-2}_N{}^J_{}
+h^N_{J+2}{}^{M'}_J \pi^{(1)}_h{}^{J+2}_N{}^J_M\right).
\nextno
$$

One can calculate the linearised momentum constraints and they are
$$
{{\pi_a^{(0)}}\over{48\pi^2 }}\left[-\half {{(n^2-1)}\over 4}
x^N_J{}^M_{} + f_J h^N_J{}^M_J\right] +{{(n^2-1)}\over 6a}
\pi^{(1)}_x{}^N_J{}^M_{} + f_J{10\over{a}} \pi^{(1)}_h{}^N_J{}^M_J=0
\nextno\eqlabel
\scalar
$$
$$
{\pi^{(0)}_a\over{48\pi^2}} h^N_{J\pm1}{}^M_J + {10\over{a}}
\pi^{(1)}_h{}^N_{J\pm 1}{}^M_J =0
\nextno\eqlabel\vector
$$
where  $f_J=\half(-1)^{2J+1}
\sqrt{(n^2-4)(n^2-1)/30}$ and $n=2J+1$.

We also have the zeroth order and linear hamiltonian constraints
$$
{1\over {8a}}{1\over{(48\pi^2)^2}} \pi^{(0)}_a a \pi^{(0)}_a +\half a^2
-{2\over 3} \Lambda a^4
=0\nextno\eqlabel\back$$
$$
\left[3a^3 \Lambda -\half a \left(n^2+\half\right)
-{1\over{(48\pi^2)^2}}{3\over{16a^2}}\pi^{(0)}_a a \pi^{(0)}_a\right]
x^N_J{}^M_{} +a f_J h^N_J{}^M_J
- {1\over{48\pi^2}}{{\pi^{(0)}_a}\over{2a^2}}\pi_x^{(1)}{}^N_J{}^M_{}=0.
\nextno\eqlabel\linear
$$
If \scalar-\linear hold then it can be shown that
$$\pi_a\; {}^v P_a = \pi_a\; {}^s P_a=0.\nextno$$
This uses the fact that $\pmatrix{1&J&J\cr a&M&{M'}\cr} c_{NN'}$ is
antisymmetric
under interchange of $(M,N)$ with $(M',N')$.

On quantisation,   the
 constraints  become operators that annihilate
the wave function $\Psi$. If a factor ordering is chosen as in
\scalar-\linear\ (i.e. just put hats on everything as it stands) then
$$\hat\pi_a\; {}^v \hat P_a\Psi = \hat \pi_a\; {}^s \hat P_a\Psi=0
\nextno$$
so
$$\hat\pi_a\; \hat P_a^{(2)}\Psi = \hat \pi_a\; {}^t \hat P_a\Psi.
\nextno
$$
Since $\hat \pi_a\Psi \ne 0$ and $\hat \pi_a$ commutes with
$\hat P^{(2)}$ and
${}^t \hat P$ this implies that
$$ \hat P_a^{(2)}\Psi = {}^t \hat P_a\Psi.
\nextno
$$

Thus the second order constraints on
the wave function may be reduced to the condition that
its dependence on the transverse traceless modes
be $SO(4)$ invariant.

\newsection{A Scalar Field}

So far we've dealt only with vacuum cosmologies. The treatment
of classical linearisation
instability  was originally confined to the vacuum case.
Results in the non-vacuum case vary according to the
problem being considered. Kastor and Traschen [\ref\traskast\TRASKAST]
investigate the
case where the perturbations in the energy-momentum
of the matter are prescribed
at some initial time, either directly or by specifying how the
constituent fields vary. This leads, in the case where the
background spacetime has ``Integral Constraint Vectors'',
to constraints on the possible metric variations allowed.
Arms [\ref\arms\ARMS] investigates the linearisation stability
of the Einstein-Maxwell equations without specifying the
matter perturbations. She finds that linearisation instability
will occur if the (spatially compact)
background space-time has Killing vectors
which  generate diffeomorphisms under which the $U(1)$ connection
is invariant. A similar calculation is done for Einstein-Yang-Mills
[\ref\armstwo\ARMSTWO].

With this in mind, suppose we want to add a massive minimally coupled
scalar field, $\Phi$, to the model, where $\Phi$ has a
background homogeneous part and an
inhomogeneous  perturbation. Now, the spatial Killing
vectors generate rotations which leave the background scalar field
invariant. Thus we expect linearisation instability to occur.
Indeed, the matter part of the momentum constraint is given by
$$ \H_m^i=g^{ij} {{\partial \Phi}\over{\partial x^j}} \pi_\Phi.
\nextno$$
It is easy to see that smearing this with a Killing vector
and calculating the first order part will give identically zero
since ${\cal{L}}_{e_a}\Phi^{(0)}={\cal{L}}_{e_a}\pi_\Phi^{(0)}=0$.

We expand $\Phi$ and $\pi_\Phi$ in scalar hyperspherical harmonics,
which are the $SU(2)$ representation matrices,
$$
\eqalignno{
\Phi(\gamma,t)& ={\phi}(t)+\sum f^{NM}_J(t) \D^J{}_{NM}(\gamma)
 \sqrt{{{J(J+1)}\over
{16\pi^2}}}\cr
\pi_\Phi(\gamma,t) &={1\over{16\pi^2}}\sqrt q \pi_{\phi}(t)
 +\sqrt q \sum
\pi_f{}^{NM}_J(\gamma) \D^J{}_{NM}(\gamma)
 \sqrt{{{J(J+1)}\over
{16\pi^2}}}&\nextnoalign\cr
}
$$
where ${\phi}(t)$ and $\pi_{\phi}$ are ``background quantities''
and the rest is
the perturbation.
Then, we see that $\H_m^{(0)}{}^i=0$ and
$$
(P_m)_a^{(1)}=a^2\int d^3x q_{ij}\H_m^{(1)}{}^i e_a{}^j=0.
\nextno$$

The second order of the matter part of the constraint is
$$
(P_m)_a^{(2)}= a^2\int d^3x \left(q_{ij}\H_m^{(2)}{}^i
+h_{ij}\H_m^{(1)}{}^i\right)e_a{}^j
\nextno
$$
which can be calculated to be
$$(P_m)_a^{(2)}= \sum_{JNM} \pmatrix{1&M&J\cr a&J&{M'}\cr}
\sqrt{J(J+1)(2J+1)}\;
f^{N}_J{}^{M'}\pi_f^{(1)}{}^J_N{}^{}_M
\nextno
$$ and similarly
$$(\tilde P_m)_a^{(2)}=
-\sum_{JNM} \pmatrix{1&N&J\cr a&J&{N'}\cr}
\sqrt{J(J+1)(2J+1)}
f^{N'}_J{}^{M}\pi_f^{(1)}{}^J_N{}^{}_M.
\nextno
$$
Thus, the $SO(4)$ invariance extends to the matter dependence
of the wave function.

\newsection {Discussion}

We have seen how linearisation instability arises in a model
of quantum cosmology in which departures from homogeneity
are treated perturbatively. It gives rise to second order
constraints on the wave function which imply that the wave function
is  $SO(4)$ invariant. This is as it should be of course
since a field configuration
on the three-sphere and a rotated configuration are the {\it{same}}
as far as quantum cosmology is concerned. Notice that wave functions
calculated in this model may be interpreted either
as wave functions of the universe or the quantum states of wormholes,
depending on the boundary conditions. This analysis
does not distinguish between them and all wave functions
 are required to  be rotationally invariant.
Similar considerations
would arise in any model of perturbations around a mini-superspace
with closed spatial sections.

We saw how linearisation instability manifested itself in a
non-vacuum model in which the background matter field was
invariant under the transformation generated by the Killing fields.
It might be possible to prove a general result along these lines.
Indeed it is conjectured [\ref\wald\WALD] that such a result would
hold for any  theory described by a  hamiltonian which takes a
``pure constraint'' form. This is supported by the linearisation
instability of the Einstein-Yang-Mills equations [\ref\arms\ARMS,
\ref\armstwo\ARMSTWO].

In Section 5, we used the zeroth order hamiltonian constraint to
show that the vector and scalar parts of the second order constraints
were redundant once the lower order constraints were imposed.
In ref.[\ref\hallhawk\HALLHAWK] it is not the zeroth order hamiltonian
constraint that is imposed on the wavefunction but the homogeneous
projection of the hamiltonian constraint. This is \back\ plus a
part which is quadratic in the perturbation. Note that while it
is not clear how this is justified in the perturbative approach,
using this homogeneous hamiltonian constraint or the zeroth
order hamiltonian constraint does not affect our result since the
difference will be a higher order than that to which we are working.

Finally, this
calculation shows how neatly the hyperspherical
spinor harmonics exploit the group structure
of $S^3$. One could use them to calculate explicitly the
action of the DeSitter group on wave functions of gravitational
perturbations on a DeSitter background. Six of the ten second
order constraints are those calculated in section 4. The remaining
four correspond to the boost Killing vectors, $B_{\alpha}^\mu$.
In the coordinate system in which the metric
is $ds^2=-dt^2 +{1\over 4}\cosh^2 t\; q_{ij}dx^i dx^j$,
$$ B_{\alpha}^\mu = (Q_{\alpha}, 4a^{-1} \dot a Q_{\alpha}^{|i})
\nextno$$
where $Q_\alpha$, $\alpha=1,\dots 4$, are the four lowest
inhomogeneous scalar
harmonics on $S^3$ i.e. $\D^{\half}{}_m{}^n$ ($Q_{n=2}$ Lifschitz),
$a=\cosh t$, and the index $i$ is raised using $q^{ij}$.
Thus, the relevant constraint arises from
the second order term in
$$
\int d^3x\left( Q_{\alpha}\H +\tanh t\; g_{ij} Q_\alpha^{|j} \H^i\right).
\nextno
$$
\vskip .5cm
{\bf{Acknowledgments}}

I should like to thank Atsushi Higuchi for helpful comments. This work
was supported in part by the U.S. Department of Energy and by
NASA Grant No. NAGW-2381 at Fermilab.
\immediate\closeout\refout
  \vskip0.2in\noindent{\bf References}\vskip0.2in
  {\obeylines\input refout }
\end